\documentclass[10pt,twoside,twocolumn,landscape,english,manuscript,aps,preprint,showpacs,prb,superscriptaddress]{revtex4}
\usepackage[T1]{fontenc}
\usepackage[latin9]{inputenc}
\usepackage{color}
\usepackage{babel}
\usepackage{amstext}
\usepackage{amssymb}
\usepackage{graphicx}
\usepackage[unicode=true,pdfusetitle,
 bookmarks=true,bookmarksnumbered=false,bookmarksopen=false,
 breaklinks=false,pdfborder={0 0 1},backref=false,colorlinks=true]
 {hyperref}
\hypersetup{
 linkcolor=blue, urlcolor=blue, citecolor=blue}

\makeatletter

\pdfpageheight\paperheight
\pdfpagewidth\paperwidth

\@ifundefined{textcolor}{}
{%
 \definecolor{BLACK}{gray}{0}
 \definecolor{WHITE}{gray}{1}
 \definecolor{RED}{rgb}{1,0,0}
 \definecolor{GREEN}{rgb}{0,1,0}
 \definecolor{BLUE}{rgb}{0,0,1}
 \definecolor{CYAN}{cmyk}{1,0,0,0}
 \definecolor{MAGENTA}{cmyk}{0,1,0,0}
 \definecolor{YELLOW}{cmyk}{0,0,1,0}
}

\makeatother

\makeatother

\begin{document}

\title{Strain dependence of antiferromagnetic interface coupling in La$_{0.7}$Sr$_{0.3}$MnO$_{3}$/SrRuO$_{3}$
superlattices}

\author{Sujit Das}

\email[Email: ]{sujitdask@gmail.com}

\affiliation{Institute for Physics, MLU Halle-Wittenberg, 06099 Halle, Germany}

\affiliation{IFW Dresden, Helmholtzstra$\beta$e 20, 01069 Dresden, Germany}

\author{Andreas Herklotz}

\affiliation{Institute for Physics, MLU Halle-Wittenberg, 06099 Halle, Germany}

\affiliation{IFW Dresden, Helmholtzstra$\beta$e 20, 01069 Dresden, Germany}

\affiliation{Oak Ridge National Lab., Oak Ridge, 37830 TN, USA}

\author{Eckhard Pippel}

\affiliation{Max Planck Institute of Microstructure Physics, Weinberg 2, 06120
Halle, Germany}

\author{Er Jia Guo}

\affiliation{Institute for Physics, MLU Halle-Wittenberg, 06099 Halle, Germany}

\affiliation{IFW Dresden, Helmholtzstra$\beta$e 20, 01069 Dresden, Germany}

\affiliation{Institute for Physics, Johannes-Gutenberg University Mainz, 55128
Mainz, Germany}

\author{Diana Rata}

\affiliation{Institute for Physics, MLU Halle-Wittenberg, 06099 Halle, Germany}

\author{Kathrin D$\ddot{o}$rr}

\email[Email: ]{kathrin.doerr@physik.uni-halle.de }

\affiliation{Institute for Physics, MLU Halle-Wittenberg, 06099 Halle, Germany}

\affiliation{IFW Dresden, Helmholtzstra$\beta$e 20, 01069 Dresden, Germany}
\begin{abstract}
We have investigated the magnetic response of La$_{0.7}$Sr$_{0.3}$MnO$_{3}$/SrRuO$_{3}$
superlattices to biaxial in-plane strain applied $in$-$situ$. Superlattices
grown on piezoelectric substrates of 0.72PbMg$_{1/3}$Nb$_{2/3}$O$_{3}$-0.28PbTiO$_{3}$(001)
(PMN-PT) show strong antiferromagnetic coupling of the two ferromagnetic
components. The coupling field of $\mu$$_{0}$$H$$_{AF}$ = 1.8
T is found to change by $\mu$$_{0}$$\triangle$$H$$_{AF}$/$\triangle\varepsilon$
$\thicksim$ -$520$ mT \%$^{-1}$ under reversible biaxial strain
($\triangle\varepsilon$) at 80 K in a {[}La$_{0.7}$Sr$_{0.3}$MnO$_{3}$(22$\textrm{\AA}$)/SrRuO$_{3}$(55$\textrm{\AA}$){]}$_{15}$
superlattice. This reveals a significant strain effect on interfacial
coupling. The applied in-plane compression enhances the ferromagnetic
order in the manganite layers which are under as-grown tensile strain,
leading to a larger net coupling of SrRuO$_{3}$ layers at the interface.
It is thus difficult to disentangle the contributions from strain-dependent
antiferromagnetic Mn-O-Ru interface coupling and Mn-O-Mn ferromagnetic
double exchange near the interface for the strength of the apparent
antiferromagnetic coupling. We discuss our results in the framework
of available models. 
\end{abstract}


\maketitle

\section{INTRODUCTION}

Magnetic order and coupling at coherent interfaces between oxides
of perovskite type have received increasing interest during the last
decade. This includes the search for phenomena already known from
metal films, e. g. exchange bias effect between a ferro- and an antiferromagnetic
layer\cite{key-1} and the interlayer coupling through non-magnetic
spacer layers responsible for giant magnetoresistance in Co/Cu/Co.\cite{key-2,key-3}
Additionally, new phenomena have been discovered reminding one of
the two-dimensional electronic states at semiconductor interfaces,
but adding the magnetic degree of freedom to electronic interface
states.\cite{key-4} The most prominent example is the conducting
electron gas at the interface between the insulators LaAlO$_{3}$
and SrTiO$_{3}$.\cite{key-5} The interface of ferromagnetic SrRuO$_{3}$
(SRO) with ferromagnetic manganites such as La$_{0.7}$Sr$_{0.3}$MnO$_{3}$
(LSMO) is in a focus of interest, because it shows an antiferromagnetic
coupling with thus far unparalleled coupling strength in oxides.\cite{key-6}
The antiferromagnetic exchange coupling at the interface leads to
antiparallel orientation of the magnetizations of thin adjacent SRO
and LSMO layers which can be sustained in a magnetic field of several
Tesla.\cite{key-6,key-7,key-8} The strong reduction of magnetic order
at LSMO surfaces or interfaces termed as \textquotedblleft{}dead layer\textquotedblright{}
in previous work\cite{key-9} seems to be weak or absent at the LSMO/SRO
interface as has been shown, e.g., in Ref. $10$. Subsequent work
showed the complexity of magnetic order arising from combination of
the antiferromagnetic interface coupling with magnetic anisotropies
of the components which are perpendicular to the film plane and strong
for SRO and in-plane and weak for LSMO on SrTiO$_{3}$($001$) substrates,
respectively. An inhomogeneous magnetization depth profile with in-plane
Ru spins near the interface and perpendicular Ru spins inside the
SRO layer has been detected by neutron reflectivity measurements.\cite{key-11}
The magnetic order at low temperatures depends heavily on the cooling
history of samples.\cite{key-12} One reason for this is the alignment
of Ru spins during cooling through $T$$_{C}$$^{SRO}$ $\sim$150
K according to the more dominant energy of either (i) the exchange
coupling to ordered Mn spins ($T$$_{C}$$^{LSMO}$ $\geq$250 K)
at the interface, or (ii) the magnetic anisotropy energy of SRO, or
(iii) the Zeeman energy in an applied magnetic field.\cite{key-12}
At low temperatures, the magnetic anisotropy of SRO is so large that
full alignment of Ru spins is hard to achieve in applied magnetic
fields of a few Tesla. Hence, the arrangement of Ru spins during cooling
is (partially) \textquotedblleft{}frozen in\textquotedblright{}. 

Meaningful investigation of magnetic coupling at oxide interfaces
has been enabled by the advance of experimental tools such as RHEED-assisted
layer-wise growth under high oxygen pressure\cite{key-13} and scanning
transmission electron microscopy (STEM). The latter allows for semi-quantitative
evaluation of chemical intermixing at interfaces by applying the high
angle annular dark field technique (HAADF). Thermal diffuse electron
scattering at high angles ( >70 mrad) is recorded with the intensity
of the localized, incoherent scattering processes proportional to
$Z$$^{2}$ ( Z denotes the atomic number). Thus the position of atom
columns or individual atoms is imaged with a brightness related to
their atomic number, usually referred as $Z$-contrast. This technique
has been employed to characterize LSMO/SRO interfaces.\cite{key-14,key-15} 

Biaxial epitaxial strain is crucial for magnetic exchange interactions
because it systematically alters bond angles and lengths.\cite{key-16}
It has been shown to strongly affect and even reverse the sign of
Mn-O-Ru interface coupling in ultrathin SrRuO$_{3}$/AMnO$_{3}$/SrRuO$_{3}$
(A = Ca or Pr) trilayers as observed by X-ray magnetic circular dichroism.\cite{key-17}
That experiment revealed the impact of strain on the magnetic coupling
by comparing trilayers grown coherently on SrTiO$_{3}$($001$) and
LaAlO$_{3}$($001$) substrates. Superlattices (SL) of LSMO/SRO could
not be grown coherently on different substrates thus far, but rather
all published work concentrates on SLs grown on TiO$_{2}$-terminated
SrTiO$_{3}$($001$). Therefore, it seems useful to attempt $in$-$situ$
strain control on such SLs using piezoelectric 0.72PbMg$_{1/3}$Nb$_{2/3}$O$_{3}$
-0.28PbTiO$_{3}$($001$) (PMN-PT) substrates.\cite{key-9,key-18}
The strain dependence of magnetic order in SRO and LSMO single films
has been investigated earlier using in-situ strain.\cite{key-19,key-20}
Those results for bulk-like films with thicknesses beyond $50$ unit
cells ($20$ nm) can help to understand the properties of ultrathin
layers in SLs, but must be considered with care because interfaces
don't matter for the magnetization of bulk-like films. We investigate
the strain dependence of the antiferromagnetic coupling in LSMO/SRO
superlattices grown on piezoelectric PMN-PT substrates and find a
large response to reversible biaxial strain. The coupling field strongly
increases upon reversible in-plane compression which releases some
of the tensile strain in the manganite layers. The observed strain-dependent
order of Mn spins at the interface is suggested to contribute to the
strain-induced change of the apparent antiferromagnetic coupling.

\section{EXPERIMENTS}

{[}$22$$\textrm{\AA}$ La$_{0.7}$Sr$_{0.3}$MnO$_{3}$(LSMO)/ $55$$\textrm{\AA}$
SrRuO$_{3}$(SRO){]}$_{15}$ superlattices (SLs) have been grown by
Pulsed Laser Deposition (PLD) with a KrF laser (wavelength $248$
nm) on ($100$)-oriented SrTiO$_{3}$ (STO) and 0.72PbMg$_{1/3}$Nb$_{2/3}$O$_{3}$-0.28PbTiO$_{3}$(PMN-PT)
substrates using stoichiometric targets of LSMO and SRO. The laser
energy density during deposition was $3$ J/cm$^{2}$ and the frequency
$3$ Hz. The SLs are grown in $0.1$ mbar of pure oxygen at $700$
$^{\circ}$C substrate temperature. After deposition, in-situ annealing
is done at $600$ mbar O$_{2}$ at $700$ $^{\circ}$C for $45$ mins.
The deposition started with a LSMO layer and ended with a SRO layer.

The SLs have been structurally characterized by X-ray diff{}raction
in a Bruker D$8$ Discover diffractometer. The microstructure of the
SLs has been investigated by high-angle annular dark field (HAADF)
imaging in a TITAN $80$-$300$ (FEI) scanning transmission electron
microscope (STEM). The chemical interdiffusion or intermixing at interfaces
was probed by an energy dispersive X-ray spectrometer (EDX) attached
to the TITAN and operating in the STEM mode. The magnetization of
the SLs has been measured in a SQUID (Superconducting Quantum Interference
Device) magnetometer. The magnetization is expressed in Bohr magnetons
per total number of pseudocubic unit cells. The piezoelectric PMN-PT
substrates are used to carry out strain-dependent measurements.\cite{key-18,key-19}
An electrical voltage is applied along the substrate normal between
the top of the SL serving as top electrode and a NiCr/Au back electrode
of the substrate. The piezoelectric strain of the substrate is transferred
to the SL layers in spite of the large total thickness.\cite{key-9,key-21}
The magnitude of the substrate strain has been measured using x-ray
diffraction at room temperature\cite{key-21}, and the temperature
dependence has been reported in Ref.$18$.

\section{RESULTS AND DISCUSSION}

\subsection{Structural characterization}

\begin{figure}
\centering{}\includegraphics[scale=0.3]{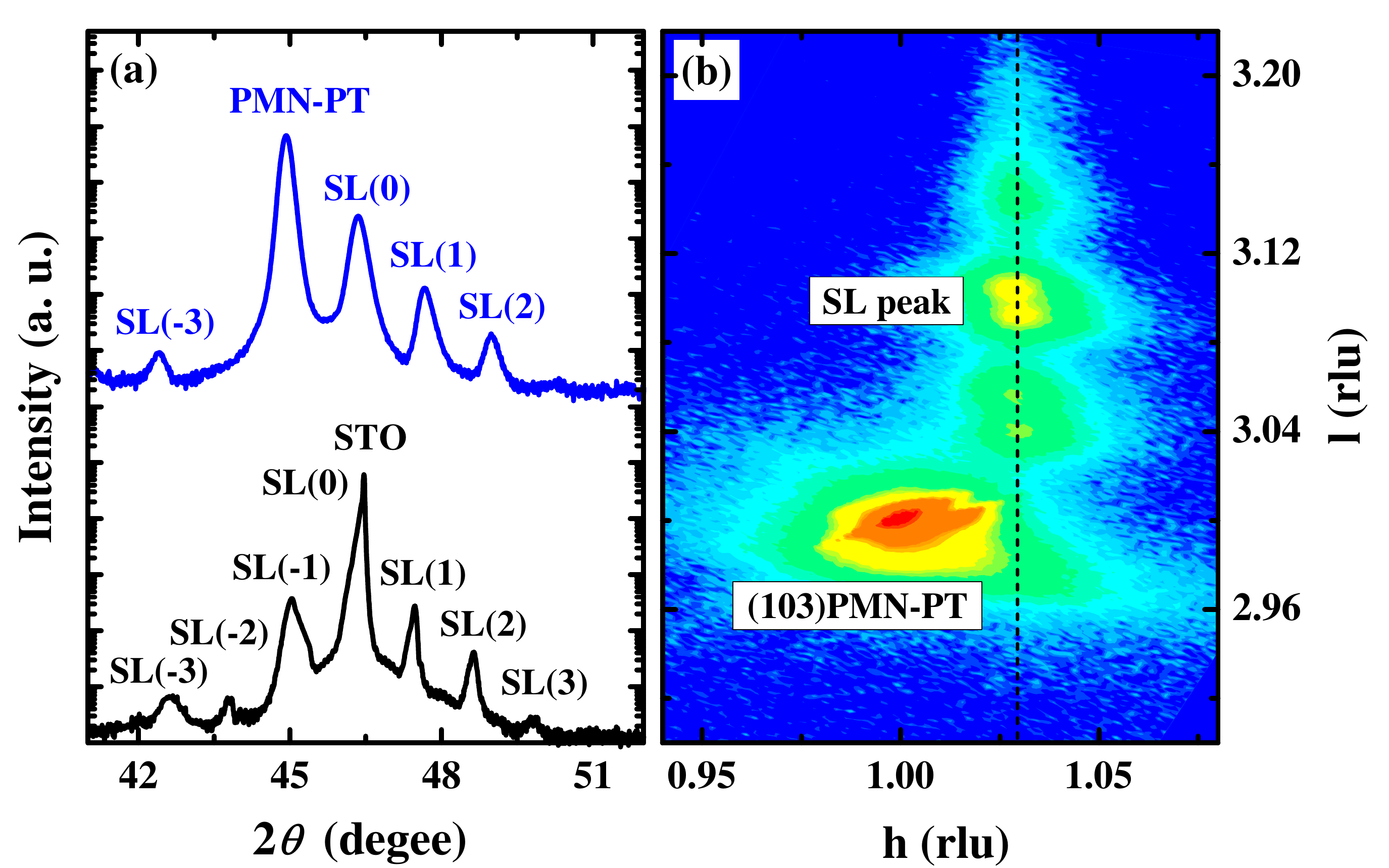} \centering\caption{(Color online)\label{fig:1} (a) $\theta-2\theta$ X-ray diffraction
scans around the ($002$) reflection of the superlattices on STO and
PMN-PT substrates, respectively. (b) Reciprocal space map around the
($103$) reflection on PMN-PT.}
\end{figure}

Fig.\ref{fig:1}(a) shows the $\theta-2\theta$ XRD scans around the
($002$) reflection of the SL grown on PMN-PT and STO, respectively.
A strong main peak and sharp satellite peaks of the SL are observed,
indicating good structural quality with sharp interfaces. The differences
in peak positions are related to the slightly different in-plane strain
of SLs on STO and PMN-PT, respectively. In order to determine the
average in-plane ($a$) and the out-of-plane ($c$) lattice parameters
of the superlattices, reciprocal space maps around the pseudocubic
($103$) reflections were recorded. The determined $c$ lattice parameters
of the SL are weighted averages over the components. According to
our XRD measurements, SLs grown on STO are strained coherently to
the substrate lattice with an in-plane parameter $a$$_{STO}$ = $3.905$
$\textrm{\AA}$. Thus, the LSMO layers in the coherently grown SL
are under tensile strain, while the SRO layers experience compressive
strain, referring to the bulk lattice parameters of $3.87$ $\textrm{\AA}$
and $3.93$ $\textrm{\AA}$ for LSMO and SRO, respectively. 

\begin{figure*}
\centering{}\centering{}\includegraphics[scale=0.45]{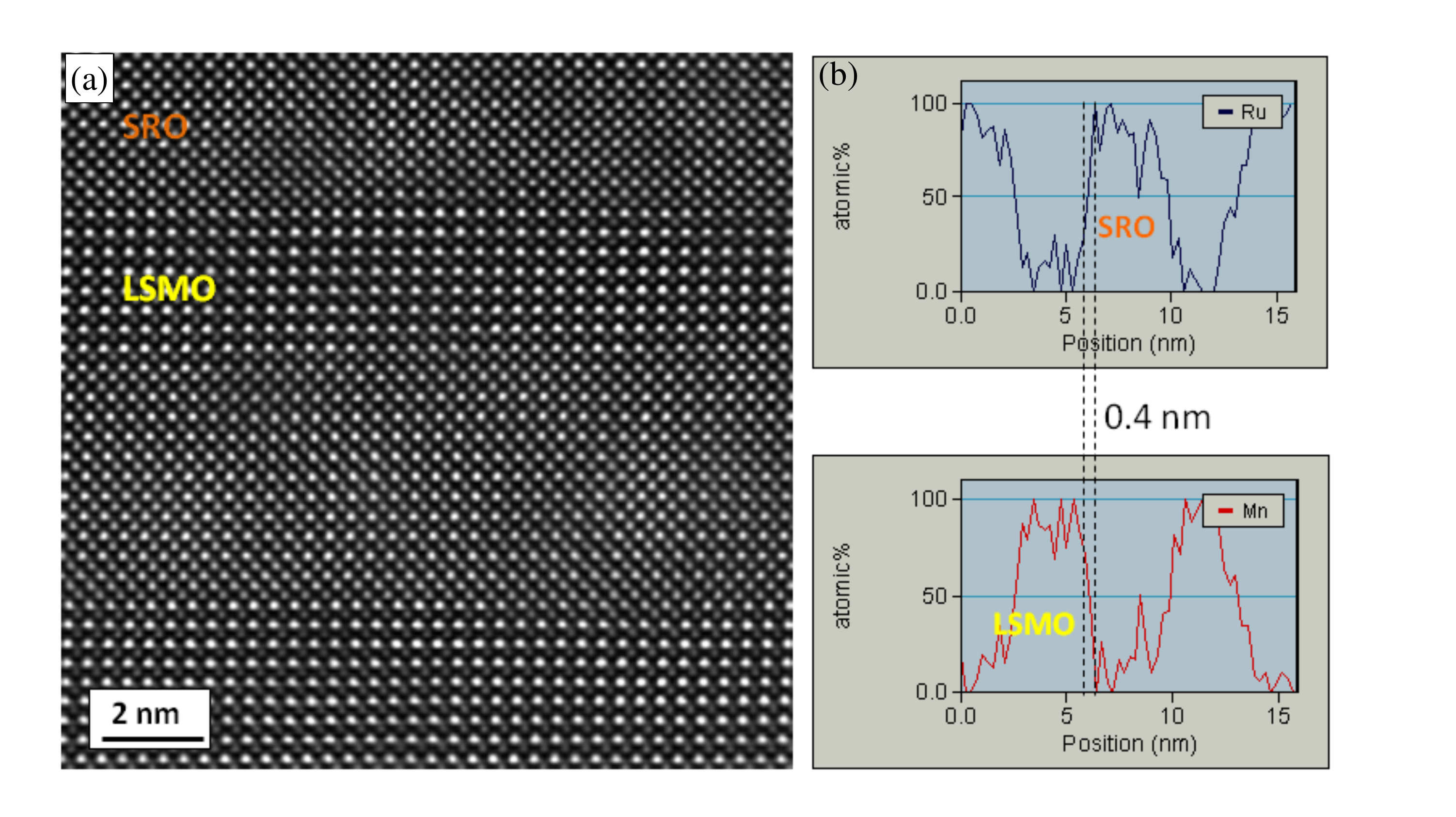}\caption{(Color online)\label{fig:2} (a) HAADF-STEM images of the investigated
SL on PMN-PT, (b) EDX line scans of Ru and Mn, crossing LSMO/SRO layers.
The dashed lines indicate an intermixing depth of about 4$\textrm{\AA}$. }
\end{figure*}

A XRD reciprocal space map of the SL on PMN-PT is shown in Fig.\ref{fig:1}(b).
The SL is not coherently strained to the PMN-PT substrate because
of the larger in-plane parameter of $a$$_{PMN-PT}$ $\backsimeq$
4.02 $\textrm{\AA}$ (which depends on ferroelectric poling). Strain
relaxation occurred immediately at the substrate-SL interface where
the first LSMO layer forms misfit dislocations. Nevertheless, the
SL itself grew coherently with a lattice parameter of $a$ = $3.92$
$\textrm{\AA}$. This has been checked by high-resolution STEM (see
below). Additionally, $in$-$situ$ recording of the in-plane parameter
by tracking the distance of RHEED diffraction streaks during growth
has been used to check for strain relaxation during growth. No strain
relaxation has been found, pointing to a coherent growth of the SL.
The in-plane lattice parameter of the SL on PMN-PT ($3.92$ $\textrm{\AA}$)
is slightly larger than that on STO ($3.905$ $\textrm{\AA}$). Hence,
LSMO layers are under slightly stronger tensile strain than in the
SL grown on STO, while the SRO layers are under very weak compressive
strain. To characterize the strain state of the components, we use
the in-plane lattice parameter and its deviation from the pseudocubic
bulk value, (whereas the out-of-plane lattice parameter of components
cannot be determined). In single layers of LSMO or SRO on STO($001$)
substrates the film structure is expected to be tetragonal (LSMO)
or orthorhombic with small monoclinic distortion (SRO), respectively,
but the symmetry of the layers in the SL might be different. For example,
it has been shown that ultrathin SRO layers in SLs with PCMO layers
are tetragonal.\cite{key-22}

High-resolution STEM images of the SL on PMN-PT confirm the absence
of dislocations and other crystal defects breaking the coherence of
the lattice inside the SL (Fig.\ref{fig:2}(a)). Probably due to the
less well-defined surface of the PMN-PT substrate (and the lattice
mismatch of the components), the SRO layers don't grow in fully flat
way, but show thickness fluctuations of $2$-$3$ unit cells. The
intermixing at the interfaces has been probed by tracking the EDX
composition along lines across the interfaces using the Ru-K$\alpha$
and the Mn-K$\alpha$ X-ray intensities (Fig.\ref{fig:2}(b)). From
this figure, intermixing of the elements Ru and Mn can be deduced
to range over a distance of about $1$ unit cell for both interfaces
LSMO/SRO and SRO/LSMO. Interestingly, intermixing is very small at
the interfaces in spite of the non-ideal flatness of the layers. This
indicates the absence of a chemical driving force for intermixing
under the applied growth conditions. No clear difference between the
interfaces of LSMO/SRO and SRO/LSMO (in the sequence of growth) has
been found, contrary to the expectation for a well-defined termination
of sharp interfaces between layers of complete perovskite unit cells.
This may result from a random termination on the PMN-PT surface or
be a consequence of the intermixing. An inspection by STEM of a SL
on SrTiO$_{3}$ substrate revealed fully coherent growth of flat layers
comparable to earlier published work by Ziese et al.\cite{key-6}
A similar magnitude of intermixing at the interfaces has been found
as for the SL on PMN-PT.

\subsection{Magnetic properties}

\begin{figure}
\centering{}\includegraphics[scale=0.295]{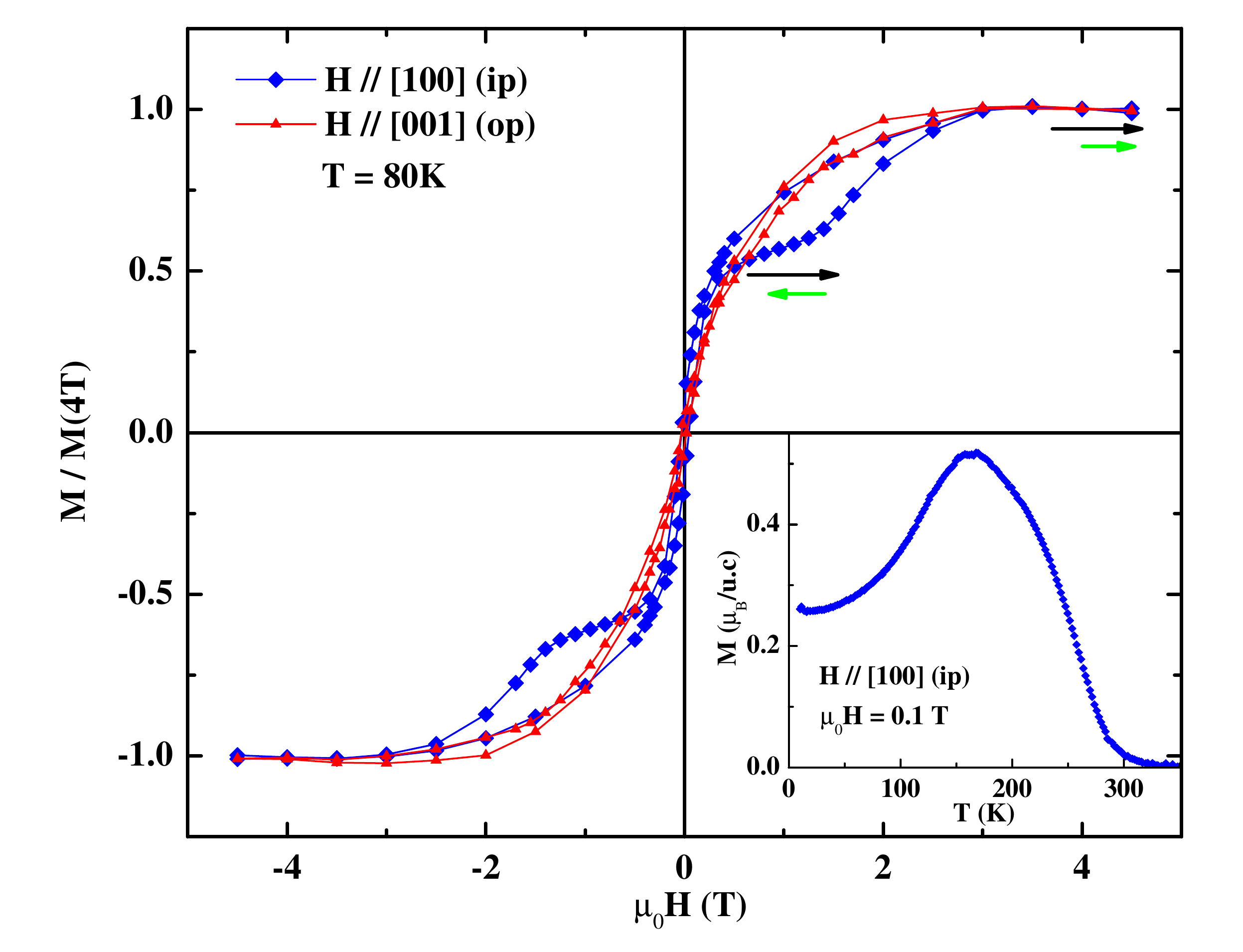}\centering\caption{(Color online)\label{fig:3} Field cooled (FC) at $2$ T in-plane
(ip) and out-of-plane (op) magnetization loops of the superlattice
on PMN-PT. Long arrow indicates LSMO magnetization and short arrow
indicates SRO magnetization. Inset: in-plane (ip) temperature dependence
magnetization at $\mu$$_{0}$H = $0.1$ T after FC the sample at
$2$ T.}
\end{figure}

We first discuss magnetization measurements of a representative SL
on PMN-PT. Temperature-dependent in-plane (parallel to an {[}$100${]}
direction) magnetization curves recorded during warming in a moderate
magnetic field such as $\mu$$_{0}$$H$ = 0.1 T after field-cooling
in $2$ T give evidence for the antiferromagnetic coupling of SRO
and LSMO layers. An example is shown in Fig.\ref{fig:3} inset where
the total magnetization is the difference of the magnetizations of
the components below the Curie temperature of SRO. The Curie temperatures
of the components, $T$$_{C}$$^{SRO}$ = 156 K and $T$$_{C}$$^{LSMO}$
= 263 K, are close to the bulk value for SRO and strongly reduced
(because of the tensile strain of $\sim$$1.3$\% and the low layer
thickness) for the LSMO layers. Magnetic hysteresis curves $M$($H$)
have been measured at temperatures between $10$ K and $100$ K both,
in the film plane along a pseudocubic {[}$100${]} direction and along
the film normal, the {[}$001${]} direction. For $T$ = $80$ K (and
in the range of $60$ \textendash{} $100$ K), $M$($H$) reveals
hard-axis behavior and nearly reversible magnetization rotation for
the normal direction (Fig.\ref{fig:3}). This result indicates spontaneous
in-plane magnetization for both layers. In-plane $M$($H$) loops
measured along a {[}$110${]} diagonal direction show smaller M($4$
T) and smaller remanent magnetization, both indicating \{$100$\}
easy axes. (In stating that, we assume biaxial in-plane symmetry not
to be broken.) 

In-plane $M$($H$) loops (Fig.\ref{fig:3}) show a two-step switching
process in the field. Firstly, the LSMO layers align along the field,
followed by the alignment of the SRO layers at 1.8 T. This switching
sequence is not immediately obvious, because strong antiferromagnetic
interlayer coupling may lead to different switching sequences depending
on the magnetic moments of both layers.\cite{key-6} Zeeman energy
in the applied field, magnetic anisotropy energy of the respective
layers and interface coupling govern the switching and may lead to
different loop shapes / switching sequences.\cite{key-23} Based on
layer thicknesses and ideal magnetization values of $3.7$ $\mu$$_{B}$/Mn
for LSMO and $1.1$ $\mu$$_{B}$/Ru one expects the magnetic moment
of LSMO layers to be larger than that of SRO layers. This would mean,
based on magnetization values, that the first switching step is related
to LSMO alignment (Fig.\ref{fig:3}), whereas the second is the SRO
alignment with the applied field. But this argumentation is weakened
by the fact that ultrathin strained LSMO layers are not fully ordered
and one does not know their magnetization well enough. More confirmation
for the switching sequence is found in the strain response as discussed
below. We assign the midpoint of the SRO transition (defined as the
point where $50$\% of the SRO magnetization has been switched) as
the coupling field $H$$_{AF}$. $H$$_{AF}$ increases from $1.4$
T to $2.8$ T when the sample is cooled from $100$ K to $10$ K.
The magnitude and temperature dependence of $H$$_{AF}$ are qualitatively
similar to earlier work on SLs on SrTiO$_{3}$($001$) substrates,\cite{key-7,key-8}
but seem to depend sensitively on the quality of the interfaces. $H$$_{AF}$
is proportional to the inverse SRO thickness,\cite{key-24} and decreases
with increased level of interface roughness / interdiffusion. There
is no information on the impact of biaxial in-plane strain on the
coupling strength available thus far. The observed strong AFM coupling
in the SL on PMN-PT indicates good structural interface quality in
agreement with the chemically sharp interfaces found by STEM. The
fluctuations in SRO layer thickness surely have the effect of broadening
the switching transition. We note that other samples prepared under
less favourable growth conditions did not show strong (or even any)
coupling; deposition parameters are vital to obtain strongly coupled
samples on PMN-PT.

\begin{figure}
\centering{}\includegraphics[scale=0.34]{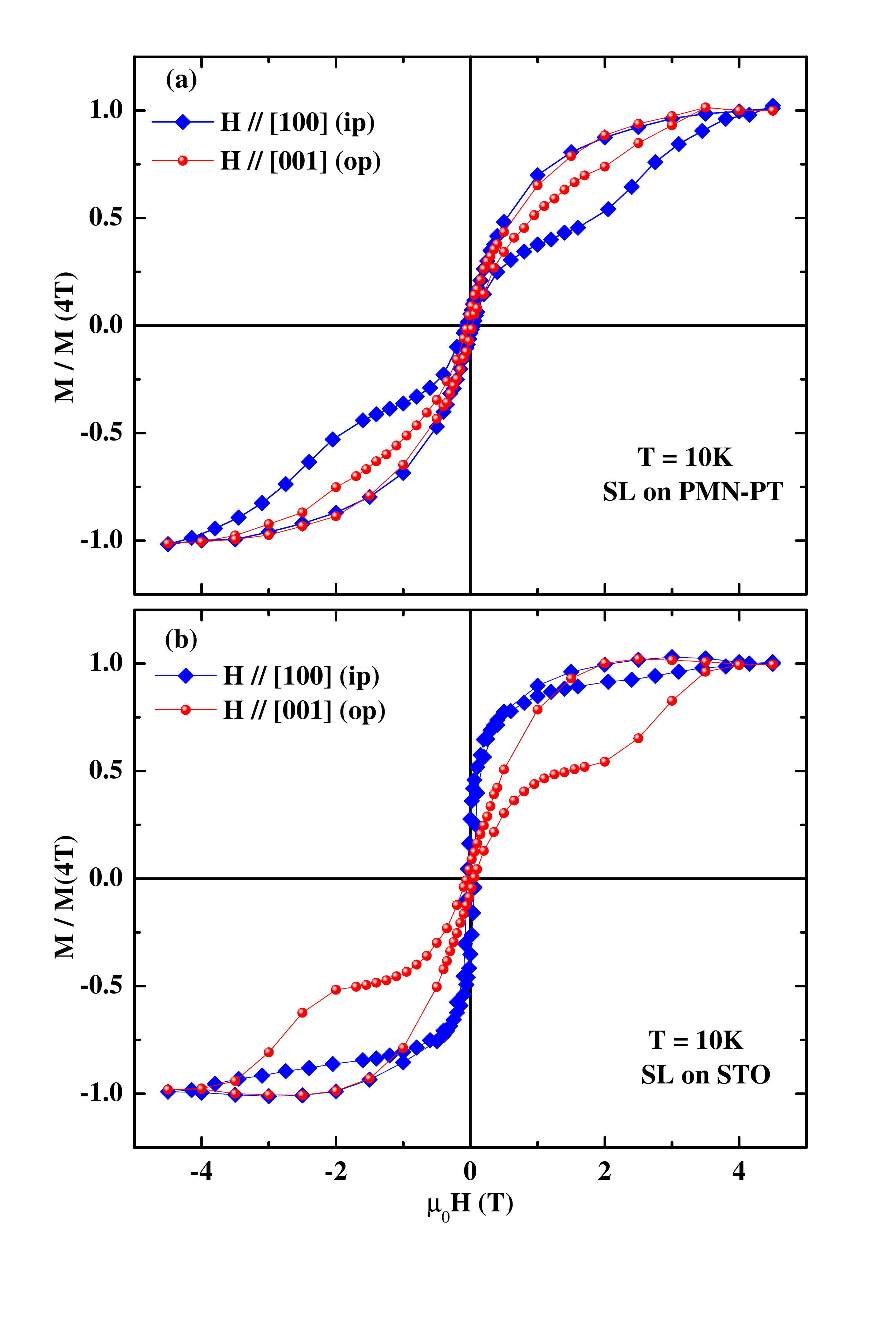}\centering\caption{(Color online)\label{fig:4} Field cooled (FC) at $2$ T in-plane
(ip) and out-of-plane (op) magnetization loops at $T$ = 10K of the
superlattices on (a) PMN-PT and (b) STO, respectively.}
\end{figure}

At $10$ K where the anisotropy of SRO is very large, the out-of-plane
magnetization is more hysteretic and reveals some remanent magnetization
(Fig.\ref{fig:4}(a)). This indicates that some SRO spins are canted
out-of-plane at $10$ K. A canted or vertical easy axis may be present
in an inner section of the SRO layers\cite{key-11} at low temperatures.
Therefore, strain-dependent measurements have been restricted to $T$
$\geq$ $60$ K where $M$ essentially lies in the film plane.

\begin{figure}
\centering{}\includegraphics[scale=0.34]{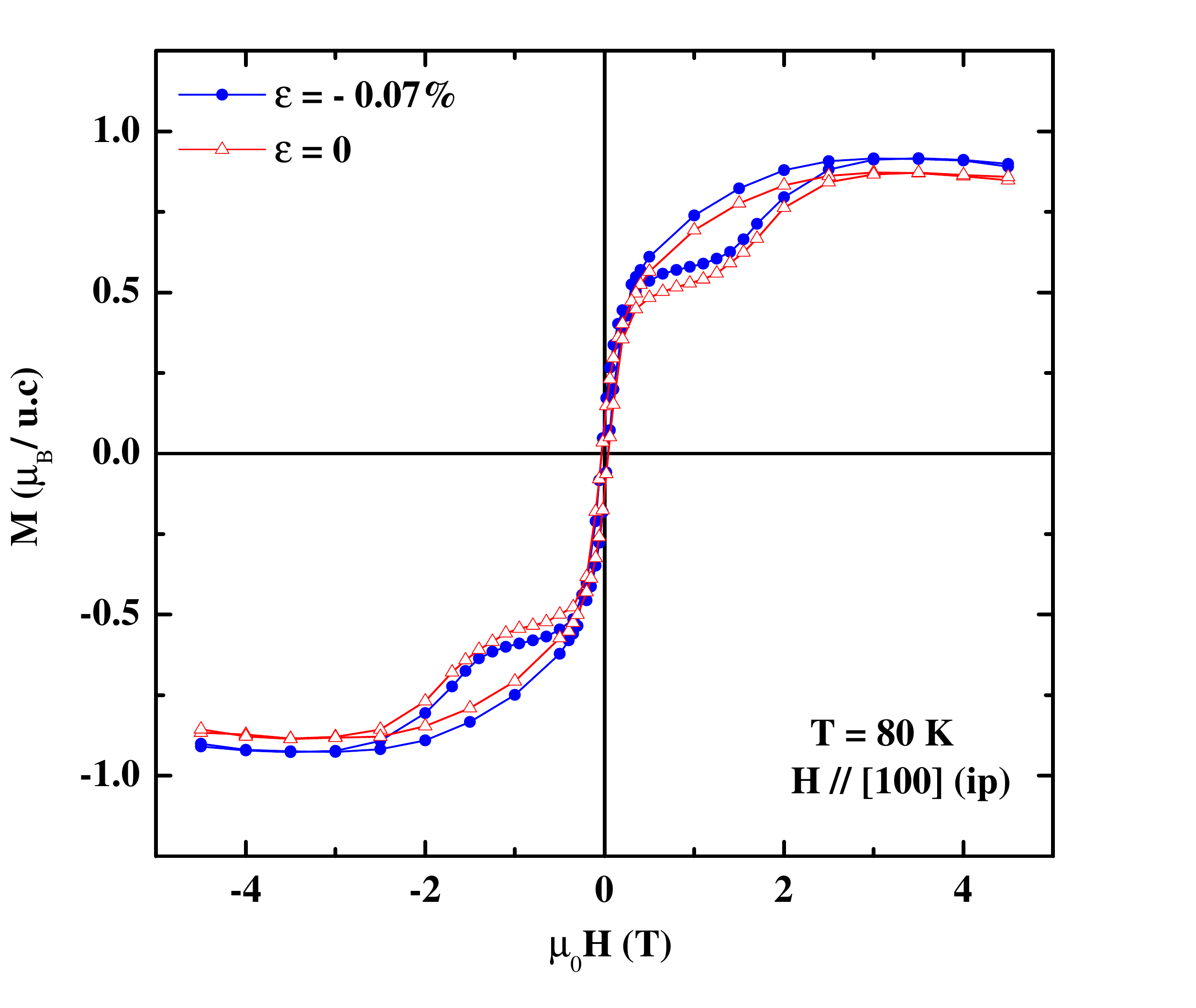}\caption{(Color online)\label{fig:5} In-plane magnetization loops of the superlattice
on PMN-PT in as-grown state ($\varepsilon$ = 0) and after piezocompression
( $\varepsilon$= - $0.07$\%).}
\end{figure}

\begin{figure}
\begin{centering}
\includegraphics[scale=0.32]{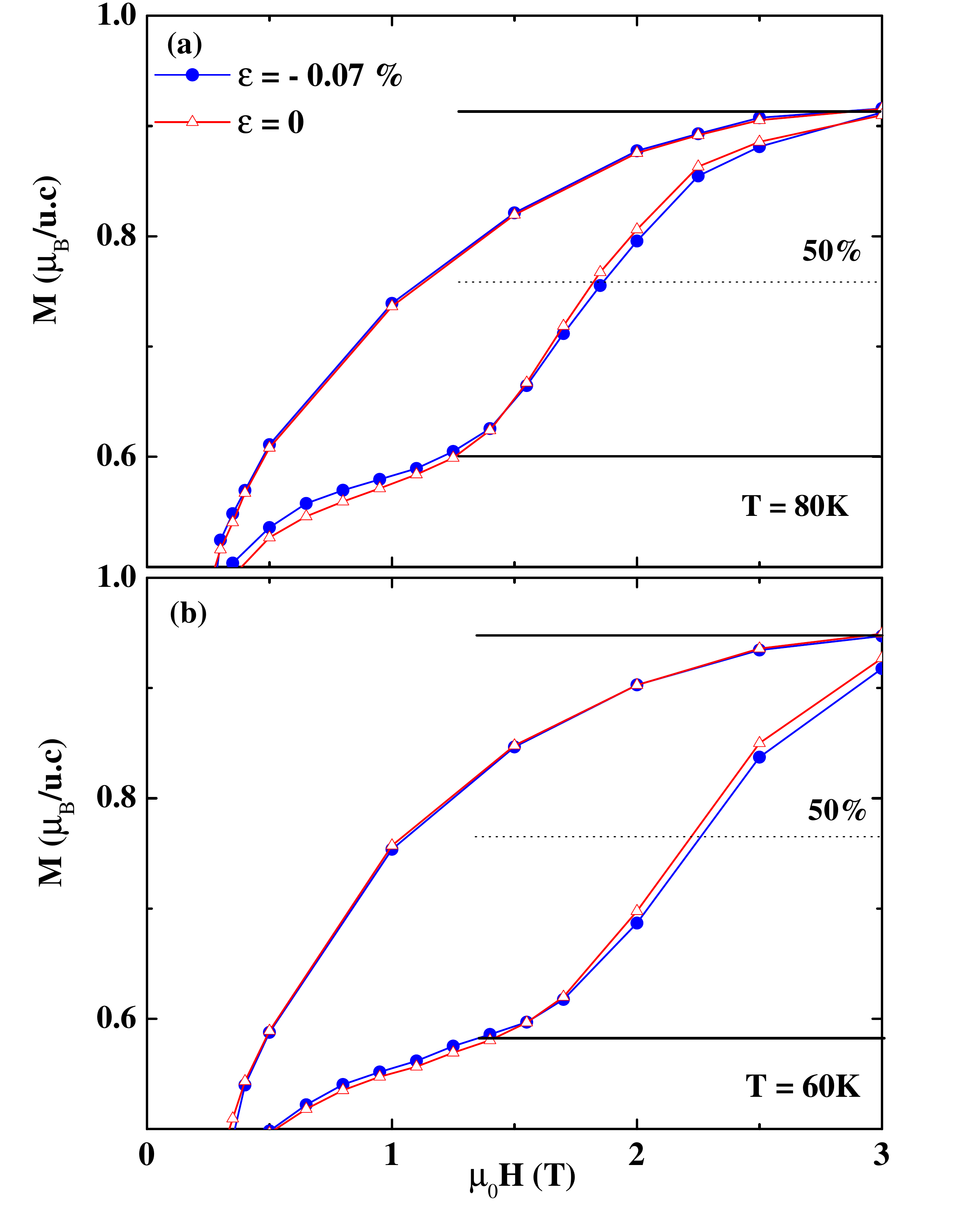}\caption{(Color online)\label{fig:6} Direct view on the change of antiferromagnetic
coupling field ( H$_{AF}$ ) induced by the piezo-compression at (a)
$T$ = 80K and (b) $T$ = 60K. We define H$_{AF}$ as the field where
50\% of the SRO magnetization has been switched.}

\par\end{centering}

\end{figure}

For inspecting the effect of biaxial strain, Fig.\ref{fig:5} gives
a comparison of the $M$($H$, $T$ = $80$ K) loops in the as-grown
and a biaxially compressed ($\triangle\varepsilon$ $\sim$ -$0.07$\%)
state. The change between the two loops is reversible and controlled
by the piezoelectric substrate strain. Similar loops have been measured
between $60$ K and $100$ K. The immediately obvious impact of the
compression is an enlargement of the saturated magnetization (at $\mu$$_{0}$$H$
= $4$ T) which roughly agrees with the enlargement seen after the
first switching step (at $\mu$$_{0}$$H$ = $1$ T) (Fig.\ref{fig:5}).
We note that the strain-induced shift of the transition field is only
visible in the expanded view in Fig.\ref{fig:6} discussed later.
Ferromagnetic order in LSMO is known to be very sensitive to tensile
strain, reflected in strong strain-induced shifts of $T$$_{C}$ for
thicker LSMO films.\cite{key-19} Ultrathin LSMO films like those
in the present SL sample show some magnetic disorder at the interfaces
which substantially reduces the LSMO magnetization. (We estimate an
ordered moment of $2.6$ $\mu$$_{B}$/Mn below.) The latter fact
makes the LSMO magnetization strain-dependent through the influence
of strain on the ferromagnetic double exchange interaction. The applied
reversible compression releases a small part of the as-grown tensile
strain of $\sim$$1.3$\% in the LSMO layers. This has a profound
effect on LSMO magnetization at $T$ <\textcompwordmark{}< $T$$_{C}$$^{LSMO}$
which increases by $6.3$\% (at $60$ K), $5.5$\% ($80$ K) or $4.4$\%
($100$ K), respectively. These values have been estimated from the
strain-induced magnetization increase observed around $1$ T (where
SRO is anti-aligned to LSMO, see Fig.\ref{fig:3}) and $4$ T (where
SRO is aligned parallel to LSMO). As expected for a strain effect
on LSMO only, the magnetization increase is the same in both cases.
This reveals a general crucial point in assessing the interlayer exchange
coupling as an independent parameter of interest, because the intralayer
magnetic order matters for the observable coupling strength. Stronger
apparent AFM coupling of the SRO layer at the interface as detected
by strain-dependent magnetization measurements may result from both,
(i) stronger Mn-O-Ru exchange interaction and (ii) higher ordered
Mn moment at the interface. (We note that the extreme case of randomly
oriented Mn moments would offer no net coupling to ferromagnetically
aligned Ru moments.) The issue is further discussed below.

Strain-induced changes of $H$$_{AF}$ have been determined as the
difference of $H$$_{AF}$ values in two investigated strain states.
Care has been taken to check the reversibility of the strain-induced
change and the reproducibility of the values in several samples and
at several temperatures. The two strain states have been measured
in immediate sequence, and curves have not been smoothed. Fig.\ref{fig:6}
provides a direct view on the change of $H$$_{AF}$ induced by the
piezo-compression in the following way: the $\varepsilon$ = 0 loop
has been shifted vertically by a constant value to match the loop
under strain at saturation ($4$ T). In this way, the strain-enhanced
LSMO magnetization is compensated. One notes the shift of $H$$_{AF}$
at the $50$\% level of the transition. The values are $\mu$$_{0}$$\triangle$$H$$_{AF}$/$\triangle\varepsilon$
= -$650$ mT \%$^{-1}$, -$520$ mT \%$^{-1}$, and -$410$ mT \%$^{-1}$
(with an error of $\sim$$20$\%) at the temperatures of $60$ K,
$80$ K, and $100$ K. (Lower temperatures have not been investigated
because the spontaneous magnetization shows some reorientation out
of the film plane as discussed above.) Further, there is a lower slope
d$M$/d$H$ of LSMO around $1$ T in the strained case. The latter
results from better ferromagnetic order of the LSMO layers after partial
release of tensile strain.

The magnetic behavior of the reference SL sample grown on STO substrate
is useful to compare because of its smaller in-plane lattice parameter.
The Curie temperatures of the components are $T$$_{C}$$^{SRO}$
= $143$ K and $T$$_{C}$$^{LSMO}$ = $305$ K. $T$$_{C}$$^{SRO}$
is not so far from the SRO bulk value, but smaller than that of the
SL on PMN-PT, in qualitative agreement with the increase of $T$$_{C}$$^{SRO}$
between a = $3.905$ $\textrm{\AA}$ and $3.92$ $\textrm{\AA}$.\cite{key-20}
$T$$_{C}$$^{LSMO}$ is about $40$ K higher on STO, an expectable
shift for the $0.4$\% weaker tensile strain of the LSMO layers. The
magnetic anisotropy of both SLs is quite different (Fig.\ref{fig:4}):
curiously, the in-plane and out-of-plane $M$($H$) loops for both
cases appear nearly like interchanged at $10$ K. Weak hysteresis
and rotation of magnetization in the field occurs for the in-plane
{[}$100$$_{pc}${]} direction on STO, whereas the out-of-plane $M$
shows a distinct transition at an antiferromagnetic coupling field
of $\mu$$_{0}$$H$$_{AF}$ = $2.8$ T. Hence, both layers of LSMO
and SRO in the SL on STO have a spontaneous perpendicular (or canted)
magnetization which is antiferromagnetically coupled. This coupling
is of similar strength like the in-plane coupling for the SL on PMN-PT.
This change of the magnetic anisotropy is consistent with the known
influence of epitaxial strain on the anisotropy in single SRO layers,
where compressed films on STO($001$) substrate show tilted perpendicular
anisotropy.\cite{key-25}

Regarding the origin of strain-dependent antiferromagnetic coupling,
we consider previously reported models. First principles calculations
by Lee $et$ $al$.\cite{key-26} reveal the lowest total energy for
the antiferromagnetic coupling of LSMO and SRO layers for an in-plane
lattice parameter close to the one we got on PMN-PT substrates. Similarly,
the antiferromagnetic state has been found in density functional theory
calculations in Ref. $6$. The influence of in-plane strain has not
been investigated yet in such calculations, to our knowledge. On the
other hand, discussion of interface magnetic coupling in oxides has
been based on orbital hybridization and strain-dependent orbital occupation
in recent work.\cite{key-17,key-27,key-28} For our lattice parameter
of $3.92$ $\textrm{\AA}$ (strong tensile strain of LSMO), Mn $e$$_{g}$
orbital energies are split leading to strong in-plane $x$$^{2}$-$y$$^{2}$
orbital occupation in Mn$^{3+}$ ions. This reduces coupling via the
$e$$_{g}$ orbitals. The piezo-compression releases a small part
of tensile strain and enhances the probability of electrons to occupy
out-of-plane orbitals ($4$d $t$$_{2g}$ xz and yz minority orbitals
for Ru, $3$d $e$$_{g}$ 3$z$$^{2}$-$r$$^{2}$ for Mn). Hence,
one would expect stronger hybridization and magnetic coupling under
piezo-compression, in line with the observed sign of the strain effect
on antiferromagnetic coupling. The details in an orbital picture seem
to be less clear if one uses previously suggested arguments. Seo et
al.\cite{key-17} have discussed a strain-dependent orbital occupation
of Ru$^{4+}$ ions at interfaces of SRO with various manganites, and
find a stronger antiferromagnetic coupling for the larger in-plane
parameter. This agrees with their experimental results (for different
manganites than LSMO), but conflicts with our observation. In a step
beyond, the contributions of $e$$_{g}$ orbitals have been considered.
In SrRuO$_{3}$, the Ru$^{4+}$ $e$$_{g}$ orbitals are empty because
of the large crystal field splitting. In Mn$^{4+}$, they are empty,
whereas in Mn$^{3+}$ there is one $e$$_{g}$$^{\uparrow}$ electron.
Nominally, LSMO contains $30$\% of Mn$^{4+}$ and $70$\% Mn$^{3+}$
ions. Coupling via the $e$$_{g}$ $3$$z$$^{2}$-$r$$^{2}$ orbitals
of Mn and Ru would thus be antiferromagnetic for Mn$^{4+}$ and ferromagnetic
for Mn$^{3+}$ at the interface according to the Goodenough-Kanamori
rules. The $e$$_{g}$ $3$$z$$^{2}$-$r$$^{2}$ orbital occupation
of Mn$^{3+}$ is expected to increase with in-plane compression, because
the single $e$$_{g}$ electron gets a higher probability to leave
the tensile-strain-stabilized $x$$^{2}$-$y$$^{2}$ orbital. Again,
this $e$$_{g}$-orbital-related mechanism reduces the total antiferromagnetic
coupling upon in-plane compression and thus disagrees with our result.
Possibly, these single-orbital considerations cannot describe the
unusually strong antiferromagnetic coupling at the LSMO/SRO interface
if it was based on itinerant electrons forming a joined band for both
components.

One more option should be considered, that is a non-ideal interface
structure. Interdiffusion of about one unit cell can strongly affect
the experimentally observable coupling. Recently, it was shown that
Mn ions of lower oxidation state can even reside on the A site of
the ABO$_{3}$ perovskite lattice in case of a strong Mn excess.\cite{key-29}
If such a situation would occur at the LSMO/SRO interface, additional
magnetic coupling pathways would be present. Such a mechanism present
at non-ideal interfaces may also influence experimental results and
calls for further improvement of knowledge on real interface structures. 

One outcome of this work is the finding that it is difficult to characterize
the Mn-O-Ru interface coupling based on magnetization measurements
if the Mn-O-Mn coupling at the interface is changing simultaneously.
This is clearly true for our experiment, as is seen in the enhanced
saturated magnetization of the LSMO layers upon piezo-compression.
Investigating interface coupling through magnetization measurements
means to take into account the intralayer magnetic order in both components
as well as the exchange coupling at the interface. Manganite layers
are known to show some degree of magnetic disorder at interfaces.
In our experiment, this is evident from the lower saturated moment
of LSMO as follows. For the as-grown state, the magnetic moment of
$\sim$$0.6$ $\mu$$_{B}$ per unit cell of the superlattice at $1$
T is assumed to represent LSMO layers aligned and SRO layers anti-aligned
with the field (Fig.\ref{fig:5}). The reversal of SRO layers yields
a change by $\sim$$0.3$ $\mu$$_{B}$ / u.c., leading to an estimated
ordered moment of $2.6$ $\mu$$_{B}$ / Mn, in contrast to $3.7$
$\mu$$_{B}$ / Mn for fully ordered Mn spins. Release of tensile
strain is known to enhance the ferromagnetic Mn-O-Mn double exchange
interaction in LSMO, in line with the observed larger LSMO magnetization
upon in-plane compression. Hence, we expect the increased antiferromagnetic
coupling of SRO layers to result partially from better ordered Mn
spins at the interfaces.

\section{CONCLUSIONS}

Summarizing, coherent superlattices of {[}LSMO($22$ $\textrm{\AA}$)/SRO($55$
$\textrm{\AA}$){]}$_{15}$ on piezoelectric PMN-PT substrates show
strong antiferromagnetic interface coupling with a profound dependence
on reversible strain. The coupling field $H$$_{AF}$ is enhanced
by $\sim$$50$ mT per $0.1$\% of reversible biaxial compression
(for a superlattice in-plane parameter of $3.92$ $\textrm{\AA}$).
Simultaneously, the magnetic order of the LSMO layers changes strongly
with the strain. We see the latter effect as an important second influence
on $H$$_{AF}$ besides the strength of the Mn-O-Ru exchange interaction;
it is possibly even dominating the observed strain effect. The strain
dependence of antiferromagnetic coupling in LSMO/SRO has not yet been
understood based on first principles theory or an orbital hybridization
scenario.

\section*{ACKNOWLEDGMENT}

This work was supported by the DFG within the Collaborative Research
Center SFB 762 \textquotedblleft{}Functionality of Oxide Interfaces.\textquotedblright{}
We thank A. Ernst for discussions.

\end{document}